\documentclass[a4paper,11pt]{article}
\pdfoutput=1 

\usepackage{jinstpub} 

\usepackage{gensymb}

\usepackage{lineno}

\title{\boldmath The ATLAS Insertable B-Layer: from construction to operation}

\author[a,1]{A. La Rosa,\note{Corresponding author.}}

\affiliation[a]{Max-Planck-Institut f\"ur Physik (Werner-Heisenberg-Institut),\\ F\"ohringer Ring 6, D-80805 M\"unchen, Germany}

\emailAdd{alessandro.larosa@cern.ch}

\abstract{The ATLAS  Insertable B-Layer (IBL) is the innermost layer of pixel detectors, and was installed in May 2014 at a radius of 3.3 cm from the beam axis, between the existing Pixel detector and a new smaller radius beam-pipe. The new detector, built to cope with high radiation and  occupancy, is the first large scale application of 3D sensors and CMOS 130\,nm technology.\\
The IBL detector construction was completed within about two years (2012\,--\,2014), and the key features and challenges met during the IBL project are presented, as well as its commissioning and operational experience at the LHC.}


\collaboration[c]{on behalf of ATLAS collaboration}

\proceeding{8$^{\text{th}}$ International Workshop on Semiconductor Pixel Detectors for Particles and Imaging\\
5-9 September 2016\\
Sestri Levante, Italy}

\begin{document}
\maketitle
\flushbottom

\section{Introduction}
ATLAS~\cite{ATLAS} is a general purpose experiment operating at the Large Hadron Collider (LHC) at CERN. The ATLAS detector was designed to be sensitive to a wide range of physics signatures to fully exploit the physics potential of the LHC at a nominal luminosity of $10^{34}$\,cm$^{-2}$s$^{-1}$. \\
The Pixel detector~\cite{PIXEL} is the innermost part of the ATLAS detector and consists of three barrel layers and three disks on each side to guarantee  at least three space points over the full tracking pseudo-rapidity range.
It was designed to operate up to  a total ionising dose (TID) of  50\,MRad, a fluence of $10^{15}$\ 1-MeV\,n$_{\mathrm{eq}}$cm$^{-2}$  and a peak luminosity of $10^{34}$\,cm$^{-2}$s$^{-1}$.\\
This paper describes the key features and challenges met during the construction and operation of a new innermost pixel detector, the Insertable B-Layer (IBL)~\cite{IBLTDR}, added during the first long shutdown of the LHC in 2013-2014 between the first pixel layer (B-Layer) and a new smaller radius beam pipe. The main motivation of the IBL is to maintain or improve the ATLAS performance during Phase-I LHC operation despite possible irreversible radiation damage effects in the B-Layer of the Pixel detector, as well as the increasing bandwidth requirements resulting from the expected Phase-I LHC peak luminosity.

\section{The ATLAS Insertable B-Layer (IBL) detector}
\subsection{Design, construction and qualification}
\begin{figure}[h!]
\centering
\includegraphics[scale=0.6]{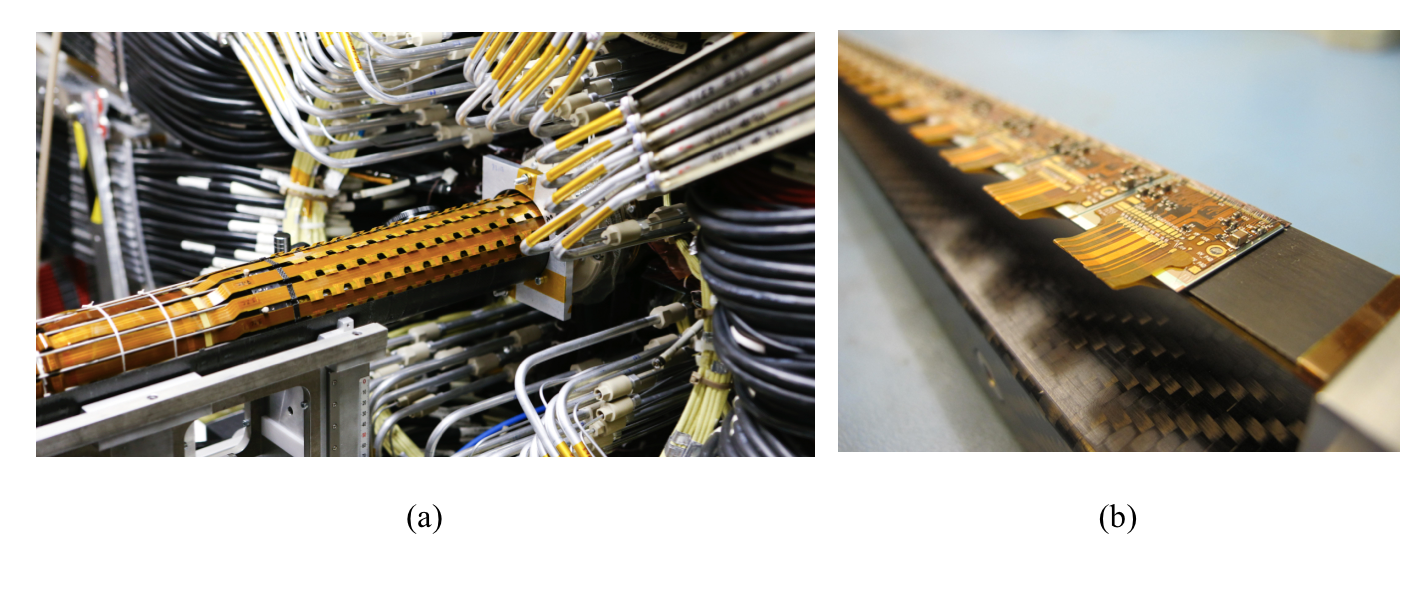}
\caption{(a) The ATLAS IBL detector prior the insertion. (b) An IBL stave where the single detector modules are mounted on carbon fibre support structures.}
\label{fig:IBL-detector-view}
\end{figure}

\begin{figure}[h!]
\centering
\includegraphics[scale=0.25]{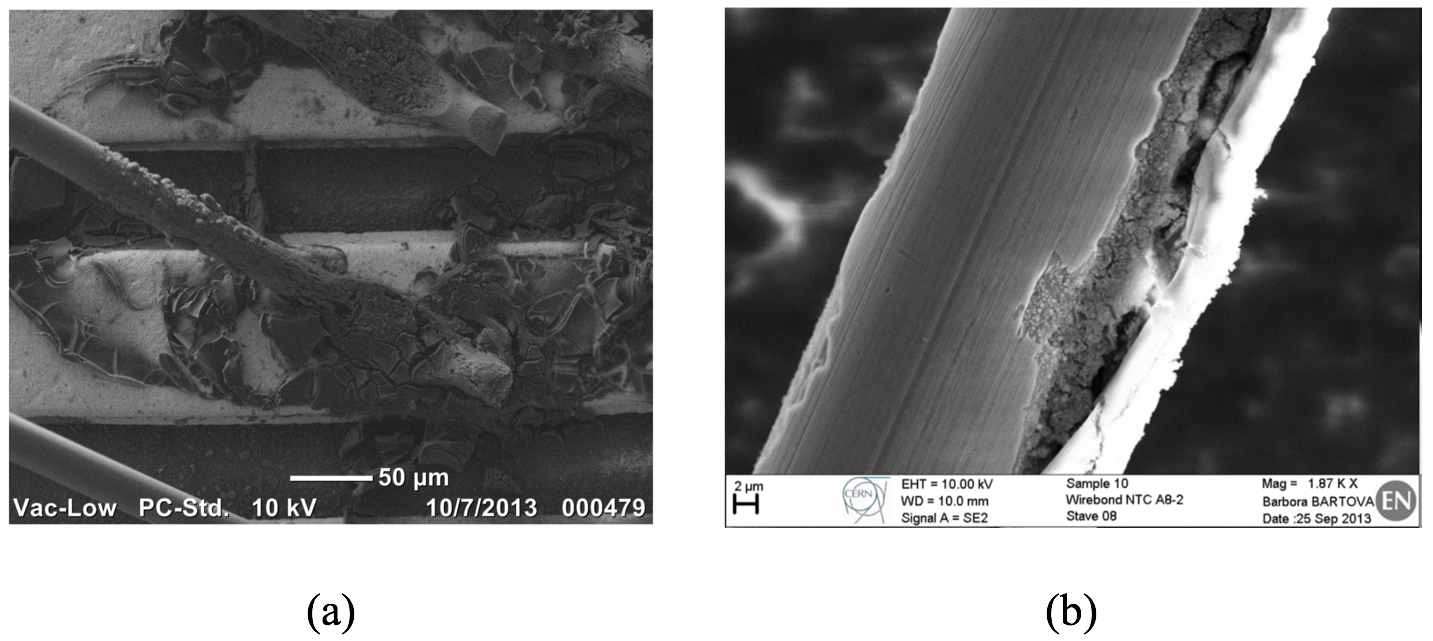}
\caption{(a) Scanning Electron Microscopy (SEM) image of corroded wires and residues and (b) image of a corroded Al-wire taken from one of the corroded stave.}
\label{fig:IBL_corrosion}
\end{figure}

The IBL detector (Figure\,\ref{fig:IBL-detector-view}) is the additional innermost pixel layer that has been built around the new beryllium beam pipe and then inserted inside the Pixel detector in the core of the ATLAS detector.  It consists of 14 carbon fibre staves each 2\,cm wide and 64\,cm long, and tilted by 14\degree in $\phi$ surrounding the beam-pipe at a mean radius of 33\,mm and covering a pseudo-rapidity of $\pm$\,3. Each stave has an integrated CO$_{2}$ cooling pipe, and is equipped with 32 FE-I4 front-end chip~\cite{FEI4} bump bonded to silicon sensors.
The FE-I4 chip is fabricated with 130\,nm CMOS technology and consists of 26880 pixel cells organised in a matrix of 80 columns (50\,$\mu$m pitch) by 336 rows (250\,$\mu$m pitch). Each front-end cell contains an independent, free running amplification stage with adjustable shaping time, followed by a discriminator with an independently adjustable threshold. The FE-I4 keeps track of the firing time of each discriminator as the time-over-threshold (ToT) with 4-bit resolution, in counts of an external supplied clock of 40\,MHz nominal frequency. A common sensor footprint for engineering and system purpose was chosen for the pixel module considering that there are two different silicon sensor technologies: planar n$^{+}$-on-n manufactured by CiS (Germany)~\cite{PLANAR} and 3D with passing through columns manufactured by FBK (Italy) and CNM (Spain)~\cite{3D}.
Twelve two-chip planar modules cover the central part of the stave while four single-chip 3D modules cover the forward regions of both ends of the stave. \\
The IBL detector was designed to reach $300~{\rm fb^{-1}}$, which will expose the detector to a fluence of 2.5\,$\times$\,10$^{15}$ 1\,MeV n$_{\mathrm{eq}}$\,cm$^{-2}$ and  a total ionising dose (TID) of 100\,Mrad.
Considering the high radiation levels expected for the IBL  detector, all of its components have been qualified up to a fluence of 5\,$\times$\,10$^{15}$ 1\,MeV n$_{\mathrm{eq}}$\,cm$^{-2}$ with a TID of 250\,Mrad.\\
The modules (around 710 unit) have been produced, mechanically and electrically qualified in 2013. 
The first batch of both module types had large numbers of bump bonding failures, which was traced back to the excessive flux in the flip-chipping process. 
The final yields of produced modules were 75\% for the planar two-chip modules, 63\% for the 3D single-chip CNM modules and 62\% for the 3D single-chip FBK modules; these yields exclude the first production batch~\cite{ALE}.
\newline
Two production staves were damaged during stave QA; some electrical dysfunctions were found when the staves were cooled to -20\,$\degree$C, caused by ice building up on the coldest part of the staves.
It was observed that most of the wire bonds were corroded and for some wire bonds were ruptured due to corrosion (Figure\,\ref{fig:IBL_corrosion}). 
Similar, less severe, damage was found on other production staves. 
Of the 12 staves already produced, 11 were affected by the corrosion; the decision was to cure (clean) and rework (rewire-bond) them.
%
%
The corrosion of the staves was caused by water accumulating on the wire bond pad of the flexes; during the thermal cycle procedure each stave was embedded in a plexiglass handling frame, to protect from any accidental damage; the temperature inside this frame was different to the temperature in the climate chamber.
This temperature difference was due to the large climate chamber volume (1.63\,m$^{2}$), and the rapid temperature increase (from -40C to 40 C in 10 minutes); consequently the area inside the frame reached dew point for several minutes during each thermal cycle, causing the build up of water on the IBL.
Electrical characterisations and metrology surveys confirmed that the corrosion did not affect the electrical or mechanical activity of the staves. The remaining staves were not thermally cycled during production.\\
The construction of the IBL detector lasted two years, and a  total of 710 FE-I4 modules were produced, 400 of which were loaded onto staves. 20 staves were produced and fully qualified, 18 of them were available for loading and the best 14 were selected for being integrated into the detector. 
The stave loading phase lasted one year with two interruptions, for the investigations of the bump-bonding failure,  and the wire-bond corrosion. 
Both issues were solved during the production and 0.1\% of pixels were dead on the 14 staves used in the IBL~\cite{QAnote}.\\
%
%
%
\subsection{Integration and commissioning}
\begin{figure}[h!]
\centering
\includegraphics[scale=0.45]{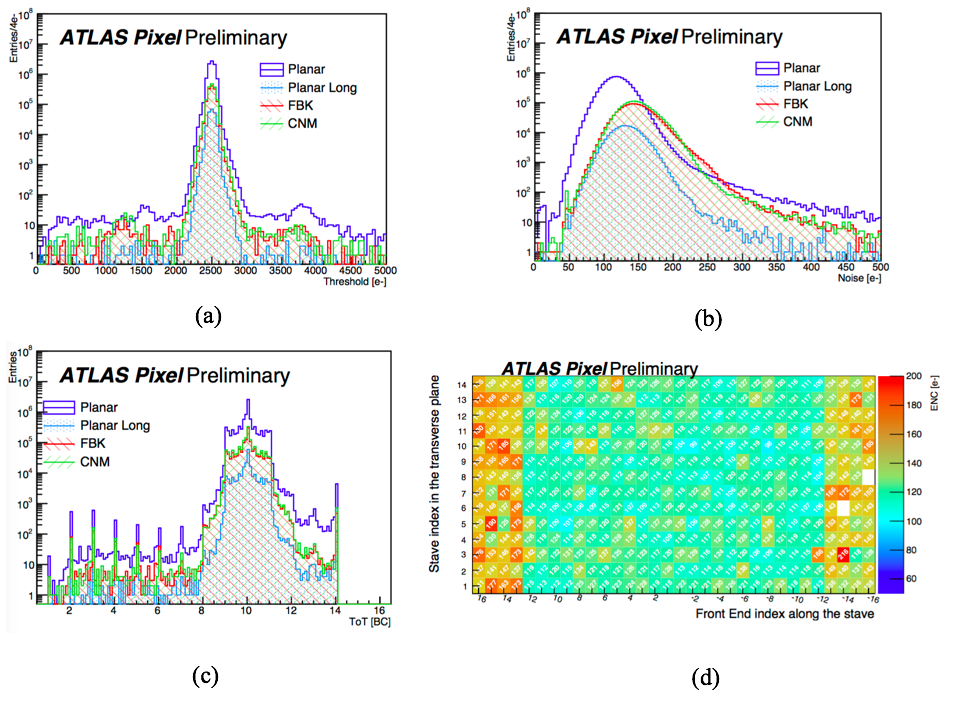}
\caption{(a) The measured threshold and (b) equivalent noise charge  for each pixel in the IBL detector, as a function of sensor type and (d) per front end, after tuning the threshold to 2500 electrons. (c) The measured mean  time-over-threshold (ToT) for each pixel in the IBL detector, as a function of  sensor type, after tuning the ToT to 10 BC (one BC = 25 ns) for 16000 electrons.}
\label{fig:IBL_calibration}
\end{figure}

In May 2014 the IBL detector was successfully installed inside ATLAS and the commissioning started. The first set of tests investigated the electrical integrity of the IBL detector right after installation. Detailed comparisons between QA results and commissioning test results confirmed that the module operation were identical before and after the integration. 
Dedicated FEs settings were required before the IBL could be added to the ATLAS data-taking. This phase was divided into two main operations: the tuning of the FE parameters related to the threshold and the time-over-threshold (as shown in Figure\,\ref{fig:IBL_calibration}.) and the timing calibration (i.e. the optimisation of the timing response for each front end pair, needed to maximise the efficiency in one single bunch crossing readout window).
%
%
\newline 
The IBL was found to have a mechanical distortion during cosmic run commissioning and alignment. This is caused by a difference in the coefficients of thermal expansion (CTE) of the  stave components. 
A 3-dimensional finite element analysis (FEA) was performed to understand the observed large distortion of the staves.  
The temperature dependence of the distortion was investigated during 2015 cosmic runs at various temperatures. Track-based alignment corrections determined the position of IBL modules, as well as their displacement relative to the nominal geometry. 
The distortion magnitude was found to depend linearly on the operating temperature of the IBL, with a gradient of around 10\,$\mu$m\,K$^{-1}$.
The maximum local displacement of the IBL module relative to the nominal geometry was found to be around 300\,$\mu$m at -20\,$\degree$C. It was possible to correct for this displacement with track-based alignment, when the distortion is static over time. 
Based on the data collected from the temperature sensors of the IBL detector during the stable cosmic runs, the size of the peak-to-peak temperature variation of the IBL staves was measured to be less than 0.2\,K. The effect of the distortion due to temperature variation of 0.2\,K was estimated, and the expected bias to the transverse impact parameter (d$_{0}$) of charged tracks is about 1\,$\mu$m, which is small in comparison to the expected d$_{0}$ resolution, $O$(10\,$\mu$m)~\cite{Distortionnote}.
\subsection{Operation}
\begin{figure}[h!]
\centering
\includegraphics[width=\textwidth]{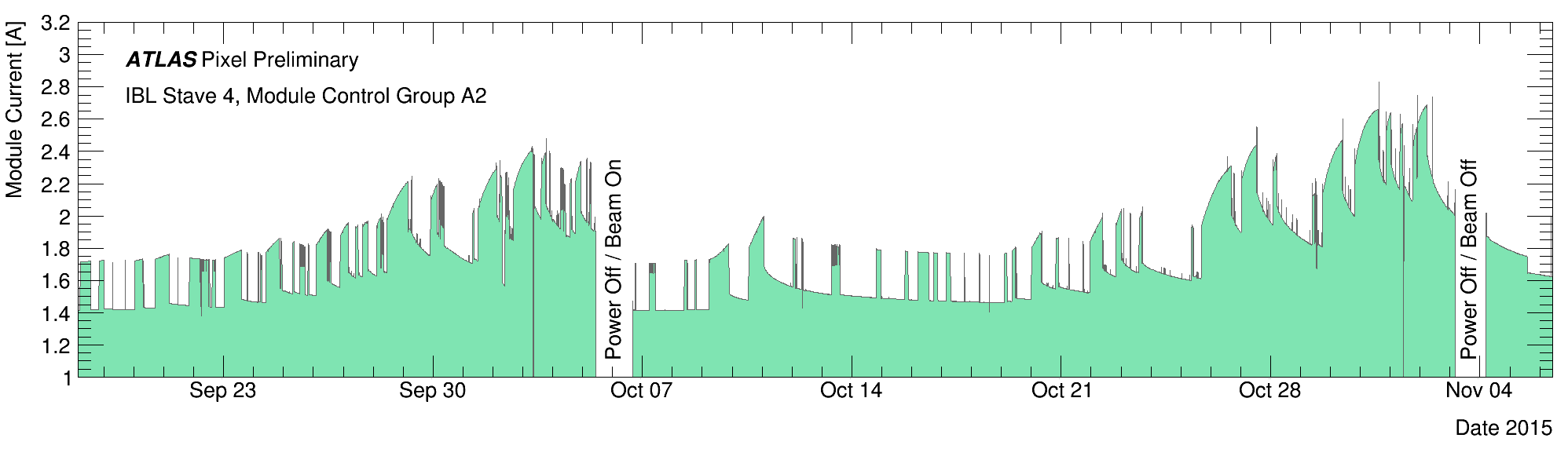}
\caption{Evolution of low-voltage current drift of four FE-I4 chips from the middle of September until the beginning of November 2015. Two levels of the current are shown depending on the configuration of the chip: STANDBY (lower level, not for data taking) and READY (higher level, for data taking).}
\label{fig:IBL_LVdrift}
\end{figure}

\begin{figure}[h!]
\centering
\includegraphics[scale=0.3]{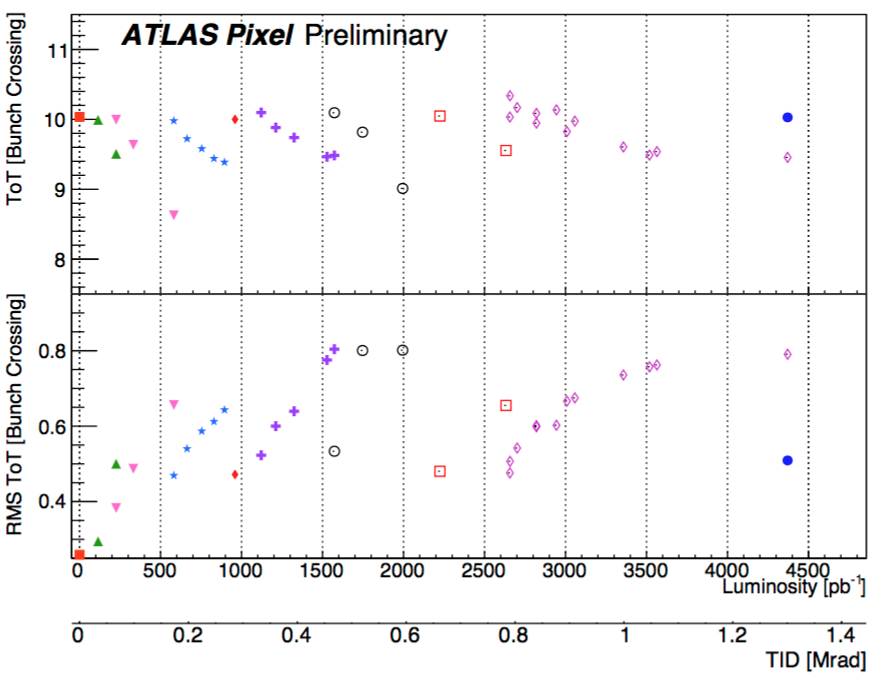}
\caption{
The evolution of the mean and RMS of the measured time-over-threshold (ToT) over all pixels in the IBL detector as a function of the integrated luminosity and the corresponding total ionising dose (TID), as measured in calibration scans. A ToT of 10\,BC (one BC = 25\,ns) corresponds to 16\,k electrons.}
\label{fig:IBL_ToTdrift}
\end{figure}

The IBL detector has been successfully operating since 2015 improving the ATLAS tracking performance~\cite{SOSHI}.\\
A significant increase of the low voltage current consumption of the front-end chip was observed during operation, along with detuning of the threshold and time-over-threshold values. These changes were due to TID.
An example of the low-voltage current drift of FE-I4 chips is shown in Figure\,\ref{fig:IBL_LVdrift}, while the evolution of the measured time-over-threshold over all pixels in the IBL detector as a function of the integrated luminosity and the corresponding TID in 2015 is showed in Figure\,\ref{fig:IBL_ToTdrift}.
Radiation effects caused the measured time-over-threshold to drift as integrated luminosity increased, but short periods of annealing and regular re-tunings brought the mean ToT values back to the optimal value. The detector was regularly retuned, and each marker type in~Figure\,\ref{fig:IBL_ToTdrift} corresponds to a single tuning of the detector.\\
\begin{figure}[h!]
\centering
\includegraphics[scale=0.45]{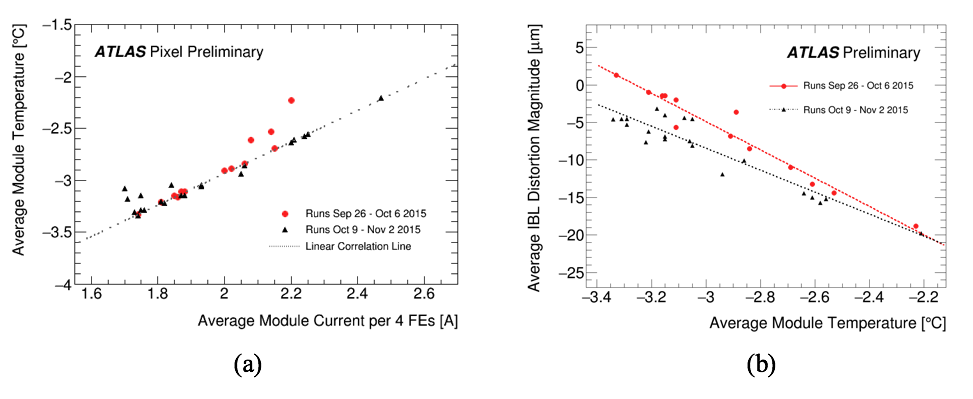}
\caption{Performance of the IBL modules during high luminosity proton-proton collision runs from September to November 2015, separated into the periods before (red circles) and after (black triangles) the long power-off on October 6. 
(a) The data are displayed as a function of the average module current per 4 front-ends of the IBL. 
(b) The data are displayed as a function of the average module temperature of the IBL \cite{Distortionnote}.} 
\label{fig:IBL_DistortionTemp}
\end{figure} 
\newline
Increasing low voltage currents heated the IBL modules, changing the module temperature and affecting the distortion magnitude  (Figure\,\ref{fig:IBL_DistortionTemp}).\\
The increase of the low voltage current  of the FE-I4 and the drifting of its operating parameters were tracked back to the generation of a leakage current  in NMOS transistors induced by radiation~\cite{FACCIO}, where
two main radiation damage mechanisms happen. 
The first is related to the trapping of radiation-induced positive charges in the shallow trench isolation (STI) oxide at the edge of the transistor, and their accumulation creates an electric field that opens a source-drain channel, allowing a leakage current to flow through. The second is related to the formation of trapping centres in the silicon-silicon oxide interface, which mainly traps negative radiation-induced charges. 
This second mechanism is slower than the first, and reduces the electric field induced by the charges trapped in the STI, reducing the leakage current. This gives origin to the so called rebound effect. 
Dedicated lab-test measurements of irradiated FE-I4 chips showed that the increase of the leakage current reaches its peak value between 1 and 3\,Mrad, and at high TID the current decreases to a value close to the pre-irradiated one. This effect has been found to be dose rate and temperature dependent.\\
Intensive studies on the dependence of the chip current on TID, dose rate and temperature found that: i) at a given dose rate, a higher temperature of the FE-I4 chip results in a lower maximum LV current increase; ii) at a given temperature, a higher dose rate results in a higher maximum LV current increase. 
Since the measurements performed at high dose rates showed that a higher temperature results in a lower current increase, it was decided to raise the IBL operation temperature from $-\,10^\circ\text{C}$ to +\,15$^\circ\text{C}$. The digital supply voltage of the chips were lowered from 1.2\,V to 1\,V to decrease the LV currents. More details of this study and the detector operation guideline are presented in Ref.~\cite{KARO}.\\
Another challenge found during operation was to avoid wire-bond failures induced by resonant vibrations. The IBL detector has more than 50000 wire-bonds that operate within the ATLAS 2\,T solenoidal magnet field. The wire-bonds undergo to a Lorentz force when a current flows through them; considering a typical IBL wire length of 2\,mm and a current of 100\,mA, a wire in a 2\,T B-field  can undergo through a maximum force of about 4\,$\times$10$^{-4}$\,N (smaller than the minimum force required to break a wire bond of about 0.1\,N), and if the current has an AC component with a frequency close to resonance frequencies, the wire starts to oscillate with high amplitude.
The oscillation can depend on: wire length and diameter, current amplitude, B-field strength and orientation angle of the wire respects to the B-field. 
Relying on oscillation amplitude and number of cycles possible failures as micro-cracks at the wire feet can happen.\\
To mimic the IBL wire-bonds operation mode a system-test has been assembled~\cite{BEATRICE}. Custom test cards have been used to bond IBL-like wires with different lengths and orientation angles to emulate the FE chips to module-flex wire bond connections.
The cards have been then placed into two poles of an electromagnet, which produced a 2\,T B-field orthogonal to the sample under test. 
Specific wire-bonds were selected on the card and a signal with an AC current amplitude between 0 and 100\,mA was sent by a waveform generator devices, while the wires were monitored with a CCD camera.\\
Thanks to the test-results of these measurements the frequencies inducing the resonant motion of the wires were identified, and a fixed frequency trigger veto (FFTV) has been implemented in the DAQ system to exclude the potentially dangerous frequencies identified in those studies.\\
The IBL FFTV consists in two mechanisms. FFTV-T(rigger) prevents fixed trigger frequencies between 2 and 40\,kHz. It monitors the time between successive triggers, incrementing a counter when consecutive periods match within a programmable period-match-tolerance. The counter is decreased when the trigger periods do not match. If the counter exceeds a configurable threshold, a 6\,ms BUSY signal is asserted, preventing new triggers from arriving. 
FFTV-B(unch) limits the maximal trigger occupancy per bunch. The algorithm splits the LHC orbit into 112 sectors, and combines several sectors to form 56\,BC (overlapping) units. If a trigger arrives in a unit, the corresponding counter is incremented, otherwise the counter for that unit is decremented. If the counter reaches a (configurable) threshold, a BUSY signal is asserted, limiting the occupancy per bunch to 50\% or 67\% depending on the decrement (1 or 2). When the LHC is fully filled with 2808 bunches, the expected dead-time from FFTV-B is 0.12\% (default settings) at a L1 trigger rate of 100\,kHz.

\section{Summary}
The first long shutdown of the LHC in 2013-2014
was an excellent opportunity to upgrade the ATLAS Pixel detector and the Insertable B-Layer (IBL) was installed and commissioned.\\
The IBL detector construction started in 2012,  with its installation into  ATLAS in May 2014, and since Spring 2015 the detector is successfully taking data.

\newpage


\end{document}